\documentclass[prd,aps,preprintnumbers,floats,floatfix,superscriptaddress,preprintnumbers,
showpacs,eqsecnum,nofootinbib,notitlepage,
twocolumn
]{revtex4-1}
\usepackage[utf8]{inputenc}
\usepackage{latexsym,array,theorem,mathrsfs,bm,float}
\usepackage{psfrag}
\usepackage{amsfonts,amsmath,amssymb,latexsym,array,afterpage,
theorem,mathrsfs,bm,float,epsfig,color,graphicx,tabularx,here,multirow}

\begin{document}
\baselineskip=12pt

\preprint{WU-AP/1802/18, YITP-18-30, IPMU18-0065}

\title{Novel matter coupling in general relativity via canonical transformation
}
\author{Katsuki \sc{Aoki}}
\email{katsuki-a12@gravity.phys.waseda.ac.jp}
\affiliation{
Department of Physics, Waseda University,
Shinjuku, Tokyo 169-8555, Japan
}
\author{Chunshan \sc{Lin}}
\email{Chunshan.Lin@fuw.edu.pl}
\affiliation{
Institute  of  Theoretical  Physics,  Faculty  of  Physics,
University  of  Warsaw,  ul.  Pasteura  5,  Warsaw,  Poland
}
\author{Shinji \sc{Mukohyama}}
\email{shinji.mukohyama@yukawa.kyoto-u.ac.jp}
\affiliation{Center for Gravitational Physics, Yukawa Institute for Theoretical Physics, Kyoto University, 606-8502, Kyoto, Japan}
\affiliation{Kavli Institute for the Physics and Mathematics of the Universe (WPI),
The University of Tokyo, 277-8583, Chiba, Japan}

\date{\today}

\begin{abstract}
 We study canonical transformations of general relativity (GR) to provide a novel matter coupling to gravity.  Although the transformed theory is equivalent to GR in vacuum, the equivalence no longer holds if a matter field minimally couples to the canonically transformed gravitational field. We find that a naive matter coupling to the transformed field leads to the appearance of an extra mode in the phase space, rendering the theory inconsistent. We then find a consistent and novel way of matter coupling: after imposing a gauge fixing condition, a matter field can minimally couple to gravity without generating an unwanted extra mode. As a result, the way matter field couples to the gravitational field determines the preferred time direction and the resultant theory has only two gravitational degrees of freedom. We also discuss the cosmological solution and linear perturbations around it, and confirm that their dynamics indeed differ from those in GR. The novel matter coupling can be used for a new framework of modified gravity theories. 
 \end{abstract}



\maketitle

\section{Introduction}
Many models of alternative theories of gravity have been proposed so far at both high-energy and low energy scales. The ultraviolet modification of gravity would be motivated by the unification of gravity and the quantum theory, while the infrared modification is devoted to solving the dark matter and the dark energy problems. Extensions of general relativity (GR) generally lead to additional degree(s) of freedom in the gravity sector. A typical example is $f(R)$ theory in which a scalar degree of freedom appears in addition to two tensor degrees of freedom. It is well-known that $f(R)$ theory can be recast in the form of a theory of a canonical scalar field with a potential after the field redefinition $\tilde{g}_{\mu\nu}=\Omega^2 g_{\mu\nu}$ called the conformal transformation~\cite{Fujii:2003pa}. The original frame is called the Jordan frame and the frame after the conformal transformation is called the Einstein frame, respectively. $f(R)$ theory is mathematically equivalent to GR with a scalar field. However, one should notice that the matter coupling to the metric tensor is different between the Jordan frame and the Einstein frame. Even if the matter coupling is minimal in the Jordan frame, the coupling becomes non-minimal in the Einstein frame due to the conformal transformation.

The conformal transformation and its generalization, the disformal transformation~\cite{Bekenstein:1992pj}, are the powerful tools to connect two different theories. The Horndeski theory, which is the most general scalar-tensor theory with the equation of motion with at most second derivatives~\cite{Horndeski:1974wa,Charmousis:2011bf,Deffayet:2011gz,Kobayashi:2011nu,Ezquiaga:2016nqo}, is transformed into the beyond Horndeski theories~\cite{Gleyzes:2014dya,Langlois:2015cwa} via the disformal transformation~\cite{Bettoni:2013diz,Zumalacarregui:2013pma,Gleyzes:2014qga,Crisostomi:2016czh,Achour:2016rkg,BenAchour:2016fzp,Ezquiaga:2017ner}. Although the transformed theory is equivalent to the original theory~\cite{Domenech:2015tca}, one should take care that the matter field couple with which metric tensor. In this sense, two theories are mathematically equivalent in vacuum but the equivalence does not hold if we introduce a matter field. The matter fields determine the preferred frame.

In the classical mechanics, the canonical transformation is a basic variable redefinition in the Hamiltonian formulation where the Hamilton's equations are invariant under the canonical transformation. Therefore, it is interesting to ask how the basic variables are transformed under the  the canonical transformation of GR and what happens if a matter field is introduced after the canonical transformation. If we do not introduce any matter fields, two theories are equivalent under the canonical transformation. However, the matter fields must break this equivalence and then determine the preferred frame of the phase space. 

In the present paper, we thus discuss the canonical transformation of GR and provide ``new'' gravitational theories by introducing a matter field after the transformation. In the Hamiltonian formulation, time and space are separately discussed. The existence of the time and space diffeomorphism invariance is seen by the existence of the first class constraints. Since the canonical transformation does not change the structure of the Hamiltonian, the first class constraints still exist after the transformation. The resultant Hamiltonian looks different from that of GR but the theory is indeed mathematically equivalent to GR in vacuum and has the same number of degrees of freedom as GR. 

Recently, the paper \cite{Lin:2017oow} provided new class of modified gravity, called minimal modified gravity theories, in which all constraints are first class and therefore the number of the gravitational degrees of freedom is the same as (or less than) that of GR. The paper \cite{Lin:2017oow} discussed a class of minimal modified gravity theories in which there is no mixed space-time derivative terms, i.e., terms containing spatial derivatives of the extrinsic curvature. One may wonder whether the canonical transformation connects GR to the minimal modified gravity theories. However, we will show that the resultant theory after the canonical transformation generally contains the mixed space-time derivative terms. The canonical transformation generates another class of minimal modified gravity theories than \cite{Lin:2017oow}.

We then discuss the matter interaction and show that a straightforward matter coupling leads to an inconsistency result: one of the first class constraints becomes second class due to the matter interaction and then one additional mode appears in the phase space. The same conclusion is suggested in the context of the minimal modified gravity theories~\cite{Carballo-Rubio:2018czn}. However, we also provide a consistent way to introduce the matter field and give consistent new gravitational theories with two gravitational degrees of freedom.

The paper is organized as follows, in the section \ref{CTofGR} we perform a canonical transformation of GR and find that extended Hamiltonian constraint and momentum constraints are still first class, as expected. In the section \ref{naivemc}, we introduce a  scalar field representing matter sector which minimally couples to canonical transformed gravity theory. An inconsistency is spotted in this scenario. A novel and consistent matter coupling is introduced in  the section \ref{consistentmc}. We discuss the cosmology in section \ref{cosmology} and finally we make summary remarks in the last section \ref{summary}.

\section{Canonical transformation of GR}\label{CTofGR}

\subsection{Hamiltonian of GR}

We start off with the Hamiltonian formulation of GR. Throughout this paper, we shall call the Einstein frame where the Hamiltonian is given by the same one as GR and the Jordan frame which the matter fields minimally couple with, respectively.
In the $3+1$ decomposition, the Einstein frame metric is given by
\begin{align}
ds_E^2=-N^2 dt^2+\Gamma_{ij}(dx^i+N^i dt)(dx^j+N^j dt)\,.
\end{align}
Introducing the canonical variables $(N,\pi_N),(N^i,\pi_i)$, and $(\Gamma_{ij}, \Pi^{ij})$,
the total Hamiltonian is
\begin{align}
H_{\rm tot}=\int dx^3 \left[ N \mathcal{H}_0+N^i \mathcal{H}_i+\lambda_N \pi_N+\lambda^i \pi_i \right] \,, \label{tot_H}
\end{align}
where
\begin{align}
\mathcal{H}_0 &:=\frac{2}{\sqrt{\Gamma}}\left( \Gamma_{i k} \Gamma_{j l} -\frac{1}{2}\Gamma_{ij} \Gamma_{kl} \right) \Pi^{ij}\Pi^{kl}
-\frac{\sqrt{\Gamma}}{2} R(\Gamma)\,, \\
\mathcal{H}_i &:= -2\sqrt{\Gamma} \Gamma_{ij} D_k \left( \frac{\Pi^{jk}}{\sqrt{\Gamma}} \right) \,.
\end{align}
and $\lambda_N,\lambda^i$ are Lagrangian multipliers where we use the Planck units $M_{\rm pl}=1/\sqrt{8\pi G}=1$.
Since $N$ and $N^i$ are just the Lagrangian multipliers, their canonical pairs $(N,\pi_N),(N^i,\pi_i)$ can be removed from the phase space by the first class constraints $\pi_N \approx 0,\pi_i \approx 0$. The remaining independent variables in the phase space are the spatial metric $\Gamma_{ij}$ and its canonical momentum $\Pi^{ij}$. Since the Hamiltonian constraint $\mathcal{H}_0 \approx 0$ and the momentum constraint $\mathcal{H}_i \approx 0 $ are first class constraints, they reduce $4\times 2$ degrees of freedom from the variables $(\Gamma_{ij},\Pi^{ij})$. As a result, the number of the degrees of freedom is 
\begin{align}
\underbrace{10 \times 2}_{(N,N^i,\Gamma_{ij},\pi_N,\pi_i,\Pi^{ij})}
-\underbrace{4 \times 2}_{\pi_N \approx 0 ,\pi_i \approx 0}
-\underbrace{4\times 2}_{\mathcal{H}_0 \approx 0, \mathcal{H}_i \approx 0}
 = 4 = 2\times 2 \,.
\end{align}
The existence of the first class constraints $\mathcal{H}_0 \approx 0$ and $\mathcal{H}_i \approx 0 $ is related to four local translation symmetries, i.e., the invariance under the time-diffeomorphism and the spatial-diffeomorphism 
\begin{align}
t\rightarrow t'(t,x^i)\,, \quad x^i \rightarrow x'{}^i(t,x^i) \,.
\end{align}

\subsection{Canonical transformation}

For simplicity, we assume that the Jordan frame metric is obtained by the canonical transformation  $(\Gamma_{ij},\Pi^{ij})\rightarrow (\gamma_{ij},\pi^{ij})$ with a simple generating functional
\begin{align}
F=-\int d^3x \sqrt{\gamma} f(\tilde{\Pi})\,, \label{trans}
\end{align}
where $f$ is an arbitrary function of $\tilde{\Pi}$ which is the scalar quantity associated with spatial coordinate transformations defined by
\begin{align}
\tilde{\Pi}:=\Pi^{ij}\gamma_{ij}/\sqrt{\gamma}\,.
\end{align}
Since the present canonical transformation does not change the lapse and the shift, $N$ and $N^i$ are still the Lagrangian multipliers after the canonical transformation.
The old canonical pairs $(\Gamma_{ij},\Pi^{ij})$ and the new canonical pairs $(\gamma_{ij},\pi^{ij})$ are related by
\begin{align}
\Gamma_{ij}&=-\frac{\delta F}{\delta \Pi^{ij}}=f'(\tilde{\Pi}) \gamma_{ij} \,, \\
\pi^{ij}&=-\frac{\delta F}{\delta \gamma_{ij}}=f'(\tilde{\Pi}) \Pi_{ij}+\frac{1}{2}\sqrt{\gamma}\gamma^{ij}\left( f(\tilde{\Pi})-f'(\tilde{\Pi}) \tilde{\Pi} \right) \,,
\end{align}
where $f'(x)=df(x)/dx$ (we will also use the notations $f''(x)=d^2 f(x)/dx^2,f'''(x)=d^3 f(x)/dx^3$). The old variables $(\Gamma_{ij},\Pi^{ij})$ are then written in terms of the new variables $(\gamma_{ij},\pi^{ij})$ as
\begin{align}
\Gamma_{ij}&=f'(\Phi) \gamma_{ij} \,, \\
\Pi^{ij}&=\frac{1}{f'(\Phi)} \left[\pi^{ij}-\frac{1}{2}\sqrt{\gamma}\left( f(\Phi)-f'(\Phi) \Phi \right) \gamma^{ij} \right]\,,
\end{align}
where $\Phi$ is a solution to
\begin{align}
\mathcal{C}:=\pi^{ij}\gamma_{ij}-\frac{\sqrt{\gamma}}{2}\left( 3f(\Phi)-f'(\Phi) \Phi \right)=0\,. \label{constraint}
\end{align}
Hereafter, we assume that 
\begin{equation}
2f'(\Phi) - f''(\Phi)\Phi \not\approx 0\,, \label{eqn:dCdPhine0}
\end{equation}
so that (\ref{constraint}) can be solved with respect to $\Phi$. Henceforth, we shall omit the function argument of $f(\Phi)$. 

Under the condition (\ref{eqn:dCdPhine0}), $\Phi$ can be written in terms of $\pi^{ij}\gamma_{ij}/\sqrt{\gamma}$ by solving $\mathcal{C}=0$ at least in principle. In practice, however, the solution to $\mathcal{C}=0$ should be given by a complicated form in general. Therefore, we instead regard $\Phi$ as independent variables satisfying the constraint \eqref{constraint} and add the canonical pair $(\Phi,\pi_{\Phi})$ in the phase space. Since $\Phi$ is a non-dynamical variable, its canonical momentum is zero:
\begin{align}
\pi_{\Phi}\approx 0\,.
\end{align}
Then, the total Hamiltonian after the canonical transformation is given by
\begin{widetext}
\begin{align}
H'_{\rm tot}=\int d^3 \mathbf{x} \left[ N \mathcal{H}'_0+N^i \mathcal{H}'_i+\lambda_N \pi_N+\lambda^i \pi_i+\lambda_{\Phi} \pi_{\Phi}+\lambda \mathcal{C} \right]
\label{tot_H_c}
\end{align}
where 
\begin{align}
\mathcal{H}'_0(\gamma_{ij},\pi^{ij},\Phi)&:=\frac{2}{f'{}^{3/2}}\left[ \frac{1}{\sqrt{\gamma}}\left(\gamma_{ik}\gamma_{jl}-\frac{1}{2}\gamma_{ij}\gamma_{kl}\right)\pi^{ij}\pi^{kl}
+\frac{\sqrt{\gamma}}{8}(f-f'\Phi)(3f+f'\Phi) \right]
\nonumber \\
&-\frac{\sqrt{\gamma}}{2}\sqrt{f'} \left[ R(\gamma)-2\frac{f''}{f'}\vec{\nabla}^2\Phi -\left( \frac{\vec{\nabla}_i \Phi }{f'} \right)^2\left( 2f'f'''-\frac{3}{2}f''{}^2 \right)
\right] \,, \label{HC} \\
\mathcal{H}'_i (\gamma_{ij},\pi^{ij} )&:=-2 \sqrt{\gamma}\gamma_{ij} \vec{\nabla}_k \left( \frac{\pi^{jk}}{\sqrt{\gamma}} \right) \,, \label{MC}
\end{align}
\end{widetext}
and $\lambda_{\Phi},\lambda$ are Lagrangian multipliers to implement the constraints $\pi_{\Phi} \approx 0 $ and $\mathcal{C} \approx 0$. We now have 22 canonical variables $( N,\pi_N,N^i, \pi_i, \gamma_{ij}, \pi^{ij},\Phi,\pi_{\Phi} )$ and 10 constraints
\begin{align}
\pi_N &\approx 0 \,, \label{c1}\\
\pi^i & \approx 0 \,, \label{c2} \\
\pi_{\Phi} & \approx 0 \,, \label{c3} \\ 
\mathcal{H}'_0 &\approx 0 \,, \label{c4} \\
\mathcal{H}'_i & \approx 0 \,, \label{c5} \\
\mathcal{C} & \approx 0 \,. \label{c6}
\end{align}

\subsection{Number of physical degrees of freedom after canonical transformation}

Let us confirm that the canonical transformation does not change the number of degrees of freedom. First, the constraints $\pi_N\approx 0, \pi^i \approx 0$ are first class constraints which remove the canonical pairs $(N,\pi_N)$ and $(N^i, \pi_i)$ from the phase space. Note that neither the Hamiltonian constraints $\mathcal{H}'_0\approx 0$ nor the momentum constraint $\mathcal{H}_i' \approx 0$ are  first class, while, some linear combinations of the constraints yield the first class constraints which we call the extended Hamiltonian constraint and the extended momentum constraint, respectively. The extended momentum constraint is given by
\begin{align}
\mathscr{H}_i :=\mathcal{H}_i'+\pi_{\Phi} \partial_i \Phi \approx 0 \,,
\end{align}
which is indeed a first class constraint:
\begin{align}
\{ \mathscr{H}_i, \pi_{\Phi} \} &\approx 0 \,, \quad \{ \mathscr{H}_i, \mathcal{H}_0' \} \approx 0 \,,
\nonumber \\
\{ \mathscr{H}_i, \mathcal{H}_j' \} &\approx  0 \,, \quad \{ \mathscr{H}_i, \mathcal{C} \} \approx 0 \,.
\end{align}
On the other hand, the extended Hamiltonian constraint is not given by a simple expression.
To obtain it we introduce a vector $\Psi^A:=(\pi_{\Phi},\mathcal{C},\mathcal{H}_0')$ and a $3\times 3$ matrix
\begin{align}
&~ \{ \Psi_A(x),\Psi_B(y)\}
\nonumber \\
&=
\begin{pmatrix}
0 & \{\pi_{\Phi}(x),\mathcal{C} (y) \} & \{ \pi_{\Phi} (x), \mathcal{H}_0' (y) \} \\
\{ \mathcal{C} (x), \pi_{\Phi}(y) \} & 0 & \{ \mathcal{C} (x), \mathcal{H}_0' (y) \} \\
\{\mathcal{H}_0' (x),\pi_{\Phi}(y) \} & \{ \mathcal{H}_0' (x),  \mathcal{C}(y) \} & \{ \mathcal{H}_0' (x), \mathcal{H}_0' (y) \} 
\end{pmatrix}
\,.
\end{align} 
If this matrix has a zero eigenvalue and the corresponding eigenvector $v^A$ such that
\begin{align}
\int d^3y \{ \Psi_A(x),\Psi_B(y)\} v^B(y) \approx 0\,, \label{eigen_vec}
\end{align}
we obtain the first class constraint
\begin{align}
\int d^3 x \Psi_A(x) v^A (x) \approx 0\,. \label{FC}
\end{align}
The equation \eqref{eigen_vec} generally yield three independent integral equations on $v^A$. When three components of \eqref{eigen_vec} are independent, all components of $v^A$ are forced to vanish and then \eqref{FC} becomes trivial. On the other hand, when three components \eqref{eigen_vec} are not independent, i.e., \eqref{eigen_vec} admits a solution of $v^A$ parameterized by an arbitrary function of space, one can obtain a local first-class constraint which eliminates a couple of local phase space degrees of freedom. Therefore, to obtain the same number of degrees of freedom as in general relativity, the eigenvector $v^A$ has to contain an arbitrary function of space. Indeed, by using
\begin{equation}
 \{\pi_{\Phi}(x),\mathcal{C} (y) \}  = \frac{\sqrt{\gamma}}{2}(2f'-f''\Phi)\delta^3(x-y)\,,
\end{equation}
the $\pi_{\Phi}$- and $\mathcal{C}$-components of \eqref{eigen_vec} can be explicitly solved to give 
\begin{align}
v^1(x) &=\frac{2}{\sqrt{\gamma}(2f'-f''\Phi)}\int d^3y  \{ \mathcal{C}(x),\mathcal{H}_0'(y) \} v^3 (y) \,, \\
v^2(x) &=-\frac{2}{\sqrt{\gamma}(2f'-f''\Phi)}\int d^3y  \{ \pi_{\Phi} (x),\mathcal{H}_0'(y) \} v^3 (y) \,, \label{eqn:sol-v1v3}
\end{align}
and then the $\mathcal{H}_0'$-component of \eqref{eigen_vec} is automatically satisfied for an arbitrary function $v^3$, meaning that the matrix $\{ \Psi_A(x),\Psi_B(y)\}$ admits the zero eigenvalue with the eigenvector parameterized by the arbitrary function $v^3$. Therefore, we obtain the extended Hamiltonian constraint
\begin{align}
\mathscr{H}_0 \approx 0 \,,
\end{align}
from the relation
\begin{align}
\int d^3 x \Psi_A v^A = \int d^3 x \mathscr{H}_0 v^3 \,.
\end{align}
Concretely, we have
\begin{widetext}
\begin{align}
 \mathscr{H}_0(x)=
 \mathcal{H}_0'(x)
  &+ 2\int d^3y\frac{
  \{ \mathcal{H}_0'(x),\mathcal{C}(y) \} \pi_{\Phi}(y) -  \{ \mathcal{H}_0'(x), \pi_{\Phi} (y) \} \mathcal{C}(y)}
  {\sqrt{\gamma(y)}[f'(\Phi(y))-f''(\Phi(y))\Phi(y)]}\,.
\end{align}
As a result, the constraints \eqref{c3}-\eqref{c6} are divided into four first class constraints
\begin{align}
\mathscr{H}_0 \approx 0 \,, \quad \mathscr{H}_i \approx 0
\,,
\end{align}
and two second class constraints
\begin{align}
\pi_{\Phi} \approx 0 \,, \quad \mathcal{C} \approx 0\,.
\end{align}
Two second class constraints remove the variables $(\Phi,\pi_{\Phi})$ from the phase space.
The number of the degrees of freedom of the canonically transformed Hamiltonian \eqref{tot_H_c} is thus given by
\begin{align}
\underbrace{11 \times 2}_{(N,N^i,\gamma_{ij},\Phi,\pi_N,\pi_i,\pi^{ij},\pi_{\Phi})}
-\underbrace{4 \times 2}_{\pi_N \approx 0 ,\pi_i \approx 0}
-\underbrace{4\times 2}_{\mathscr{H}_0 \approx 0, \mathscr{H}_i \approx 0}
-\underbrace{2}_{\pi_{\Phi}\approx 0, \mathcal{C}\approx 0}
 = 4 = 2\times 2 \,,
\end{align}
\end{widetext}
which is the same number of the degrees of freedom as the original Hamiltonian \eqref{tot_H}.

The same conclusion is obtained by considering the consistency relations
\begin{align}
\dot{\mathcal{H}}_0'&=\{ \mathcal{H}_0' , H_{\rm tot} \}\approx 0 \,, \\
\dot{\mathcal{H}}_i'&=\{ \mathcal{H}_i' , H_{\rm tot} \}\approx 0  \,, \\
\dot{\pi}_{\Phi}&=\{ \pi_{\Phi} , H_{\rm tot} \}\approx 0  \,, \\
\dot{\mathcal{C}}&=\{ \mathcal{C}, H_{\rm tot} \}\approx 0  \,.
\end{align}
The last two equations determine the Lagrangian multipliers $\lambda_{\Phi}$ and $\lambda$. Substituting them into the first two equations, we find the consistency relations are weakly satisfied which means $N$ and $N^i$ are undetermined. Hence, there must be four first class constraints associated with $N$ and $N^i$ which are indeed the extended Hamiltonian constraint and the extended momentum constraint, respectively.

\section{Inconsistency of naive matter coupling}\label{naivemc}

The Hamiltonian \eqref{tot_H_c} is equivalent to \eqref{tot_H} via the canonical transformation in vacuum. However, when we introduce a matter field minimally coupling with the Jordan-frame metric
\begin{align}
ds_J^2&=g^J_{\mu\nu}dx^{\mu}dx^{\nu}
\nonumber \\
&=-N^2dt^2+\gamma_{ij}(dx^i+N^i dt)(dx^j + N^j dt)\,,
\end{align}
the equivalence between two Hamiltonians is no longer true. The Hamiltonian \eqref{tot_H_c} with matter fields may yield a new gravitational theory.
For simplicity, we assume a minimal scalar field with a potential whose canonical variables are denoted by $(\chi,\pi_{\chi})$. The Hamiltonian is then given by
\begin{align}
\hat{H}_{\rm tot}
&=\int d^3 \mathbf{x} \Big[ N \hat{\mathcal{H}}_0+N^i \hat{\mathcal{H}}_i+\lambda_N \pi_N+\lambda^i \pi_i
\nonumber \\
& \qquad \qquad \quad +\lambda_{\Phi} \pi_{\Phi}+\lambda \mathcal{C} \Big] \label{tot_H_matter}
\end{align}
with
\begin{align}
\hat{\mathcal{H}}_0 &:=\mathcal{H}_0'+\mathcal{H}_0^{\rm m} \,, \\
\hat{\mathcal{H}}_i &:=\mathcal{H}_i'+\mathcal{H}_i^{\rm m} \,,
\end{align}
where $\mathcal{H}_0'$ and $\mathcal{H}_i'$ are given by Eqs. \eqref{HC} and \eqref{MC}, and the matter parts are given by
\begin{align}
\mathcal{H}_0^{\rm m} &=\frac{1}{2\sqrt{\gamma}}\pi_{\chi}^2+\sqrt{\gamma} \left( \frac{1}{2}(\partial_i \chi)^2+V \right) \,, \\
\mathcal{H}_i^{\rm m} &=\pi_{\chi} \partial_i \chi \,.
\end{align}
Even though adding a matter field, $\pi_N\approx 0, \pi^i \approx 0$ and the extended momentum constraint
\begin{align}
\hat{\mathscr{H}}_i := \hat{\mathcal{H}}_i'+\pi_{\Phi} \partial_i \Phi \approx 0 \,, \label{EMC}
\end{align}
are the first class constraints. However, we find that the matrix $\{\hat{\Psi}_A(x), \hat{\Psi}_B(y) \}$ does not have an zero eigenvalue where $\hat{\Psi}_A=\{ \pi_{\Phi}, \mathcal{C}, \hat{\mathcal{H}}_0 \}$. Indeed, the $\pi_{\Phi}$- and $\mathcal{C}$-components of \eqref{eigen_vec} (with the new set $\hat{\Psi}_A$) can be explicitly solved to give \eqref{eqn:sol-v1v3} with $\mathcal{H}_0'$ replaced by $\hat{\mathcal{H}}_0$, and substituting them into the $\hat{\mathcal{H}}_0$-component results in inconsistency in general. For example, for the simplest case $f=\alpha \Phi$ we have 
\begin{align}
&~ \int d^3y \{ \Psi_3 (x),\Psi_B(y)\} v^B(y) 
\nonumber \\
\approx& 
(1-\alpha^{-1})\left[ (\pi_{\chi}\vec{\nabla}^2 \chi+\vec{\nabla}_i \chi \vec{\nabla}^i \pi_{\chi}) v^3 + 2 \pi_{\chi} \vec{\nabla}_i \chi \vec{\nabla}^i v^3 \right] \,, \label{eq_v3}
\end{align}
which does not vanish unless $\alpha=1$. Therefore the matrix $\{\hat{\Psi}_A(x), \hat{\Psi}_B(y) \}$ has no zero eigenvalue due to the matter field if $\alpha \neq 1$ where $\alpha=1$ means the identical transformation $\Gamma_{ij}=\gamma_{ij}, \Pi^{ij}=\pi^{ij}$. As a result, the Hamiltonian \eqref{tot_H_matter} generally has seven first class constrains
\begin{align}
\pi_N\approx 0\,, \quad \pi^i \approx 0 \,, \quad \hat{\mathscr{H}}_i \approx 0 \,,
\end{align}
and three second class constraints
\begin{align}
\pi_{\Phi} \approx 0 \,, \quad \mathcal{C} \approx 0 \,, \quad \hat{\mathcal{H}}_0 \approx 0\,.
\end{align}
The number of the gravitational degree of freedom of the Hamiltonian \eqref{tot_H_matter} is $5$ in the phase space (or $2.5$ in the real space). There is an additional mode in the phase space. 

The appearance of an additional mode can be understood by that the spacetime-diffeomorphism invariance is now reduced into the invariance under the time-reparameterization and the spatial-diffeomorphism,
\begin{align}
t \rightarrow t'(t)\,, \quad x^i \rightarrow x'{}^i (t,x^i)\,. \label{time_rep}
\end{align}
Since the canonical transformation \eqref{trans} does not break the spatial-diffeomorphism invariance, the resultant Hamiltonian has the  spatial-diffeomorphism invariance which is also confirmed by that the extended momentum constrain is first class. On the other hand, the time-diffeomorphism invariance is broken by the matter coupling after the canonical transformation although there still exists the time-reparameterization symmetry. The absence of the time-diffeomorphism invariance leads to the appearance of the additional mode which is the same situation as the wrong non-projectable extensions of Horava-Lifshitz gravity~\cite{Henneaux:2009zb}.

In addition, the similar result is obtained by the paper~\cite{Carballo-Rubio:2018czn} in the context of the minimal modified gravity theories~\cite{Lin:2017oow}. The minimal modified gravity theories provide a new framework of the gravitational theories with two or less local degrees of freedom. The paper \cite{Carballo-Rubio:2018czn} pointed out that, although some of the minimal modified gravity theories admit four local first class constraints in vacuum, one of them becomes second class due to the matter coupling. This is exactly the same as the present case.

We note that the additional mode does not appear for the homogeneous configuration $\Phi=\Phi(t),\chi=\chi(t),\pi_{\chi}=\pi_{\chi}(t)$. Indeed, when $\Phi=\Phi(t),\chi=\chi(t),\pi_{\chi}=\pi_{\chi}(t)$, the equation \eqref{eq_v3} vanishes weakly and then the matrix $\{\hat{\Psi}_A, \hat{\Psi}_B \}$ has an zero eigenvalue (this is true for any function of $f$). This is due to nothing but the existence of the time-reparameterization symmetry.

\section{A consistent way of matter coupling}\label{consistentmc}

As shown in the previous section, the matter coupling leads to that the extended Hamiltonian constraint is no longer the first class constraint which can be interpreted as the matter field partially fix the time direction. The additional mode arises due to the incompleteness of the gauge fixing. One may resolve this inconsistency by fixing the gauge before introducing a matter field.
In other words, we shall introduce the gauge condition in order that the first class constraint associated with the time diffeomorphism invariance splits into a couple of second class constraints, which remain second class after introduction of matter
fields. 

In vacuum, the Hamiltonian \eqref{tot_H_c} has the first class extended Hamiltonian constraint which is the generator of the time-diffeomorphism. Hence, the gauge can be a priori fixed in order that the spacetime-diffeomorphism invariance is reduced into the invariance under \eqref{time_rep} in vacuum. The gauge fixed Lagrangian is given by 
\begin{align}
H^{\rm GF}_{\rm tot}&=\int d^3 x \Big[ N \mathcal{H}_0'+N^i \mathcal{H}_i'+\lambda_N \pi_N+\lambda^i \pi_i
\nonumber \\
&\qquad \qquad \quad +\lambda_{\Phi} \pi_{\Phi}+\lambda \mathcal{C}  + \lambda_G \mathcal{G} \Big] \,,
\end{align}
where $\lambda_G$ is the Lagrangian multipliers to implement the gauge condition
\begin{align}
\mathcal{G} \approx 0 \,, \label{gauge_fix}
\end{align}
which makes the extended Hamiltonian constraint the second class constraint, i.e.,
\begin{align}
\{ \mathcal{G}, \mathscr{H}_0 \} \not\approx 0 \,.
\end{align}
This is just a gauge fixing. However, this gauge condition turns to be ``physical'' if we introduce the matter field in the Jordan-frame after the gauge fixing:
\begin{align}
\hat{H}^{\rm GF}_{\rm tot}&=\int d^3 x \Big[ N \hat{\mathcal{H}}_0+N^i \hat{\mathcal{H}}_i+\lambda_N \pi_N+\lambda^i \pi_i
\nonumber \\
&\qquad \qquad \quad +\lambda_{\Phi} \pi_{\Phi}+\lambda \mathcal{C}  + \lambda_G \mathcal{G} \Big]  \,. \label{H_GF}
\end{align}
In this case, the equation \eqref{gauge_fix} is a physical constraint to eliminate the additional mode. The resultant Hamiltonian \eqref{H_GF} is equivalent to GR with a gauge condition \eqref{gauge_fix} via the canonical transformation in vacuum but it is not equivalent to GR when a matter field is introduced. In this formalism, the way matter field couples to the metric determines the preferred time direction.

We then discuss the condition of $\mathcal{G}$ in order that \eqref{H_GF} represent a healthy gravitational theory.
We shall retain the constraints \eqref{c1},~\eqref{c2}, and \eqref{EMC} being the first class constraints\footnote{One can generalize the case when $\mathcal{G}$ contains $N$. In this case, $\pi_N \approx 0$ is no longer the first class constraint and then the lapse is not recognized as the Lagrangian multiplier. The consistency relation $\dot{\mathcal{G}} \approx 0$ may give a constraint equation on $N$. Then, we obtain four second class constraints $\pi_N \approx 0, \mathscr{H}_0 \approx 0, \mathcal{G} \approx 0, \dot{\mathcal{G}} \approx 0$ and find the correct number of the degrees of freedom.}. Hence, $\mathcal{G}$ has to satisfy 
\begin{align}
\{ \mathcal{G},\pi_N \} \approx 0 \,, \quad \{ \mathcal{G}, \pi_i \} \approx 0 \,, \quad \{ \mathcal{G}, \mathscr{H}_i \} \approx 0\,.
\end{align}
The first two condition is trivially satisfied when $\mathcal{G}$ does not contain $N$ and $N^i$ and the third condition means $\mathcal{G}$ is a scalar density associated with the spatial-diffeomorphism. Since $\mathcal{G}$ is introduced in vacuum, it must be a function of $\gamma_{ij},\pi^{ij},\Phi$, and the time $t$.\footnote{Even if $\mathcal{G}$ is a function of $\pi_N,\pi_i,\pi_{\Phi},\pi^{ij}\gamma_{ij}$, their dependence can vanish by redefinitions of the Lagrangian multipliers $\lambda_N,\lambda^i,\lambda_{\Phi},\lambda$.}

We then take the Legendre transformation in order to obtain the Lagrangian corresponding to the Hamiltonian \eqref{H_GF}. For this purpose, we redefine the Lagrange multipliers as
\begin{align}
\lambda \rightarrow N \lambda 
\,, \quad
\lambda_G \rightarrow N \lambda_G
\end{align}
and redefine the gauge condition $\mathcal{G}$ to be a scalar quantity
\begin{align}
\mathcal{G} \rightarrow \sqrt{\gamma} \mathcal{G}
\end{align}
Just for simplicity, we assume the gauge condition $\mathcal{G}$ does not contain $\pi_{ij}$. We obtain
\begin{align}
\dot{\gamma}_{ij}&=\{ \gamma_{ij}, H_{\rm tot}^{\rm GF} \}
\nonumber \\
&= 2\vec{\nabla}_{(i}N_{j)}+\frac{4N}{f'{}^{3/2}\sqrt{\gamma}}\left( \pi_{ij}-\frac{1}{2}\pi^k{}_k \gamma_{ij} \right)+N\lambda \gamma_{ij}\,.
\end{align}
and then the conjugate momentum
\begin{align}
\pi^{ij}=\frac{\sqrt{\gamma}}{2} f'{}^{3/2}\left( K_{ij}-K \gamma_{ij}+\lambda \gamma_{ij} \right)\,,
\end{align}
where $K_{ij}=(\dot{\gamma}_{ij}-2\vec{\nabla}_{(i}N_{j)})/2N$ is the extrinsic curvature. After the Legendre transformation, the gravitational part of the Lagrangian is given by
\begin{widetext}
\begin{align}
\sqrt{-g^J}\mathcal{L}&=\dot{\gamma}_{ij}\pi^{ij}-\mathcal{H}_{\rm tot}^{\rm GF}
\nonumber \\
&= N \sqrt{\gamma} f'{}^{3/2} \Biggl[ \frac{1}{2}\left( K_{ij}K^{ij}-K^2 + \frac{R(\gamma)}{f'} \right)
-\frac{f''}{f'{}^2}\vec{\nabla}^2\Phi -\left( \frac{\vec{\nabla}_i \Phi }{f'{}^{3/2}} \right)^2\left( f'f'''-\frac{3}{4}f''{}^2 \right)
\nonumber \\
&\qquad \qquad \qquad
-  \frac{1}{4f'{}^3}(f-f'\Phi)(3f+f'\Phi)-\frac{3}{4}\lambda^2+\lambda\left(K+\frac{1}{2f'{}^{3/2}}(3f-f'\Phi) \right) -f'{}^{-3/2}\lambda_G \mathcal{G})\Biggl]\,. \label{action}
\end{align}
The variation with respect to $\lambda$ yields the equation to determine $\lambda$. Substituting it into the action, we obtain
\begin{align}
\sqrt{-g^J}\mathcal{L}&= N \sqrt{\gamma} \Biggl[ \frac{f'{}^{3/2}}{2} \left(  K^{ij}K_{ij}-\frac{1}{3}K^2 +\frac{R}{f'} \right)
+K \left( f-\frac{1}{3}f'\Phi \right)
\nonumber \\
&\qquad \qquad \qquad \quad
-\frac{f''}{f'{}^{1/2}}\vec{\nabla}^2\Phi -\left( \frac{\vec{\nabla}_i \Phi }{f'{}^{3/4}} \right)^2 \left(f'f'''-\frac{3}{4}f''{}^2 \right)+\frac{1}{3}f'{}^{1/2}\Phi^2 
-\lambda_G \mathcal{G}
\Biggl]\,.
\label{action_with_Phi}
\end{align}
\end{widetext}
The variable $\Phi$ is non-dynamical and thus it can be integrated out at least in principle. In practice, however, it is often more convenient to keep $\Phi$ as an auxiliary field in the Lagrangian. In the identical transformation case $f=\Phi$, the variation with respect to $\Phi$ yields 
\begin{align}
\Phi=-K+ {\rm terms~from~gauge~fixing~term}\,,
\end{align}
and then one can obtain the standard Einstein-Hilbert action with a gauge condition. On the other hand, in general case, the solution of $\Phi$ may not be obtained explicitly. Nonetheless, we notice that the solution is schematically expressed by $\Phi=\Phi(K_{ij},R,N)$ and thus the action after the integrating out $\Phi$ must contain the spatial derivatives of the extrinsic curvature. This mixed space-time terms have not been discussed in \cite{Lin:2017oow}. Therefore, the canonical transformations of GR connect a new class of minimal modified gravity theories.

We shall give an example of the gauge fixing.
In scalar-tensor theories, it is common to use the so-called unitary gauge in which a scalar field $\phi$ depends on only the time, $\phi=\phi(t)$. To implement an analogue of the unitary gauge in our setup, we choose $\mathcal{G}$ as
\begin{align}
\mathcal{G}:=\vec{\nabla}^2 \Phi \,,
\end{align}
which yields to a uniform variable $\Phi=\Phi(t)$, provided a proper spatial boundary condition.
In GR case $f=\Phi$, this gauge condition corresponds to the uniform Hubble slicing $K=K(t)$.
The Poisson bracket is
\begin{widetext}
\begin{align}
\{ \mathcal{G}, \mathscr{H}_0 \} 
\approx \frac{2\sqrt{\gamma} f'{}^{1/2}}{2f'-\Phi f''}
\left[ (\Phi^2 +R) \vec{\nabla}^2 +\vec{\nabla}^2 R +2 \vec{\nabla}^i R \vec{\nabla}_i 
-\vec{\nabla}^2 \vec{\nabla}^2 \right] \delta^{(3)}(x-y)\,,
\end{align}
\end{widetext}
thus, the constraints $\mathcal{G}\approx 0, \mathscr{H}_0 \approx 0$ are now regarded as two second class constraints which reduce one of the physical degrees of freedom. The time direction is fixed by $\mathcal{G}\approx 0$ and then the spacetime-diffeomorphism invariance is reduced into the invariance under \eqref{time_rep}.

After adding a matter field, $\mathcal{G}\approx 0$ turns to be a physical condition on the theory and then the time-diffeomorphism invariance is explicitly broken. Nonetheless, the time-diffeomorphism invariance can be recovered if we introduce a St\"{u}eckelberg field by promoting $t$ to a field of time and space.

\section{Cosmology}\label{cosmology}

In this section, we briefly discuss the background dynamics of the universe and the linear perturbations around the background. The gauge condition $\mathcal{G}\approx 0$ does not affect the dynamics of the FLRW spacetime because $\mathcal{G}\approx 0$ must be ``trivial'' due to the unbroken  spatial diffeomorphism invariance. Therefore, the dynamics of FLRW spacetime can be discussed without specifying $\mathcal{G}$.
For simplicity, we consider the flat FLRW universe
\begin{align}
N=\bar{N}(t)\,, \quad N^i=0 \,, \quad \gamma_{ij}=a^2 \delta_{ij}\,.
\end{align}
We find two relevant equations: 
\begin{align}
H=\frac{\Phi(4f'+\Phi f'')}{6f'{}^{1/2}(\Phi f''-2f')}\,, \label{H_Phi}
\end{align}
and
\begin{align}
\frac{1}{3}\Phi^2 f'{}^{1/2}=\rho_{\rm m}\,, \label{eq_Fri}
\end{align}
where $H=\dot{a}/(a\bar{N})$ and $\rho_m$ is the energy density of a matter field. In principle, the first equation gives a solution $\Phi=\Phi(H)$. Substituting it to the second one, we obtain the Friedmann equation. For instance, the identical transformation case $f=\Phi$ gives $H=-\Phi/3$ and then Eq.~\eqref{eq_Fri} is reduced to the usual Friedmann equation $3H^2=\rho_{\rm m}$.

We then derive the quadratic action for the cosmological perturbations. The tensor mode perturbations $h_{ij}$, which is transverse-traceless part of $\gamma_{ij}$, is the gauge invariant quantities. Therefore, $\mathcal{G}$ does not need to contain $h_{ij}$. In the following, for simplicity we assume that $\mathcal{G}$ does not depend on $h_{ij}$ up to second order so that we can discuss the tensor perturbations without any additional assumptions about $\mathcal{G}$. The perturbed metric is given by
\begin{align}
N&=\bar{N}(t)\,, \quad N^i=0\,, 
\nonumber \\
\gamma_{ij}&=a^2 e^{h_{ij}}=a^2\left( \delta_{ij}+h_{ij}+\frac{1}{2}h_{ik}h^k_j+\cdots \right)
\,.
\end{align} 
The quadratic action of the tensor mode perturbations is given by
\begin{align}
S^{(2)}_h=\int dt d^3 x \frac{\bar{N}a^3f'{}^{3/2}}{8} \left[ \left( \frac{\dot{h}_{ij}}{\bar{N}} \right)^2 -\frac{k^2}{f' a^2} h_{ij}^2 \right]
\,,
\end{align}
where we have used in the momentum space and $k$ is the comoving momentum.
There is neither ghost nor gradient instabilities as long as $f'>0$ and the speed of the gravitational wave differs from unity if $f'\neq 1$. 

The vector perturbations do not have $\mathcal{G}$-dependence since the gauge mode of the vector perturbations is generated by the infinitesimal change of the spatial coordinates. The condition $\mathcal{G}=0$ is obtained by fixing the time coordinate; thus, it should not affect the vector perturbations. We shall not discuss the vector perturbations furthermore because there is no dynamical gravitational degrees of freedom in the vector perturbations. On the other hand, the scalar perturbation  may depend on a specific choice of $\mathcal{G}$ although $\mathcal{G}$ does not affect the high energy behavior of the scalar perturbations. Since $\mathcal{G}=0$ is no longer the gauge choice after introducing a matter field as for the theories with $f'\neq 1$, an inappropriate function of $\mathcal{G}$ may be problematic.

For simplicity, we assume \eqref{gauge_fix} and discuss a massless scalar field $\chi$ minimally coupling with the Jordan frame metric; that is, $\mathcal{L}_{\chi}=-\frac{1}{2}g^{J \mu\nu}\partial_{\mu}\chi \partial_{\nu}\chi$. By using the spatial gauge freedom we can assume
\begin{align}
N&=\bar{N}(1+\alpha) \,, \quad N^i=\bar{N} \delta^{ij}\partial_j \beta \,, 
\nonumber \\
\gamma_{ij}&=a^2 e^{2\zeta} \delta_{ij}\,, \quad \chi=\bar{\chi}(t)+\delta\chi
\,,
\end{align}
where the off-diagonal components of $\gamma_{ij}$ are eliminated by spatial coordinate transformation.
Substituting them into the Lagrangian, one can find that $\alpha,\beta$ and $\zeta$ are non-dynamical variables and then they can be integrated out. Finally, we obtain the quadratic order action for $\delta \chi$ as
\begin{widetext}
\begin{align}
S^{(2)}_{\chi}&=\int  dt d^3x \bar{N} a^3 
 \left[ \frac{3(k^2+a^2\Phi^2)^2}{2k^2(3k^2-a^2\Phi^2)} \left(\frac{\dot{\delta \chi}}{\bar{N}} \right)^2
-\frac{k^2}{2a^2}\left(1+\frac{a^2 \Phi^2}{k^2 f'} \right) \delta \chi^2 \right] \,.
\end{align}
\end{widetext}
We notice that the standard action for the massless scalar field is recovered in the high energy limit $k\rightarrow \infty$ as we expected. One may worry about a ghost mode appears in the low energy limit $k^2/a^2 \ll \Phi^2$. However, the low energy ghost is not problematic and it does not lead a catastrophic instability~\cite{Gumrukcuoglu:2016jbh}. Indeed, the theory with $f=\Phi$ is equivalent to GR but we know that the massless scalar field does not have any catastrophic instability in the low energy limit. In GR case $f=\Phi$, the instability of $\delta \chi$ is due to the inappropriate choice of the variable. When we define a new variable
\begin{align}
\delta X :=\delta \chi -\frac{\dot{\bar{\chi}}}{\bar{N}H} \zeta \,,
\end{align}
the action for $\delta X$ is given by the form
\begin{align}
S^{(2)}_{X}&=\int  dt d^3x \bar{N} a^3 
\nonumber \\
&\times \left[ \frac{A_1(\Phi;k)}{2} \left(\frac{\dot{\delta X}}{\bar{N}} \right)^2 - \frac{k^2 A_2(\Phi;k) }{2a^2} \delta X^2 \right]
\,.
\end{align}
The GR case $f'=1$ yields $A_1=A_2=1$ and thus the variable $\delta X$ has no instability even in the low energy limit. 

We shall discuss the leading order correction to GR. When we assume the time reflection symmetry, $f'$ has to be an even function of $\Phi$. Therefore, $f'$ may be expanded as 
\begin{align}
f'=1+\epsilon \frac{\Phi^2}{M^2}+\mathcal{O}(M^{-4})
\end{align}
where $M$ represent the scale beyond which the deviation from GR appears and $\epsilon=\pm 1$. As for the background dynamics, we obtain
\begin{align}
\Phi=-3H+\epsilon \frac{27H^3}{M^2} +\mathcal{O}(M^{-4})
\end{align}
and then the Friedmann equation is given by
\begin{align}
3H^2 \left(1-\epsilon \frac{27H^2}{2M^2} +\mathcal{O}(M^{-4}) \right)=\rho_m\,.
\end{align} 
For the massless scalar field, the coefficients of the quadratic order action are
\begin{align}
A_1&=1+ \epsilon \frac{729H^4}{2M^2(k^2/a^2+9H^2)} +\mathcal{O}(M^{-4}) 
\,, \nonumber \\
A_2&=1- \epsilon \frac{243H^4( k^2/a^2-37H^2-162H^4 a^2/k^2)}{2M^2(k^2/a^2+9H^2)^2} 
\nonumber \\
& +\mathcal{O}(M^{-4})\,.
\end{align}
The result indicates that the dynamics of the universe and the perturbations is indeed changed from the case of GR when $H\gtrsim M$ (or $\rho \gtrsim M^2$). The canonical transformations of GR generate new theories of gravity without introducing any catastrophic instability. 

We give an interesting example of the generating functional of the canonical transformation
\begin{align}
f'=(1+\Phi^4/M^4)^{-1} \,.
\end{align}
In this case, the equation \eqref{eq_Fri} indicates that the matter energy density is constrained to be the finite value $M^2/3$. As $\rho_{\rm m} \rightarrow M^2/3$, Eq.~\eqref{eq_Fri} yields $\Phi \rightarrow - \infty$ while Eq.~\eqref{H_Phi} leads to $H\rightarrow 0$ in this limit. Therefore, this theory does not have the initial singularity of the universe. Instead, the universe approaches the Minkowski spacetime as the matter energy density increases. We however note that the tensor mode perturbations are suffered from the strong coupling problem because of $f' \rightarrow 0$ in the early universe and then the perturbation theory breaks down.

\section{Summary}\label{summary}

In the present paper, we have investigated canonical transformations of general relativity (GR) to generate ``new'' theories of gravity. We first confirmed that a canonical transformation does not change the constraint algebra of the theory and thus the transformed theory has only two gravitational degrees of freedom. We then discussed the matter coupling and found a novel and consistent way of the coupling although a naive coupling leads to an inconsistent result. The matter field is introduced as follows: We first introduce the gauge fixing condition $\mathcal{G} \approx 0$ before introducing the matter field in order to reduce one first class constraint $\mathscr{H}_0\approx 0$ to two second class constraints $\mathscr{H}_0 \approx 0,\mathcal{G} \approx 0$ and then we introduce the matter field. As a result, the matter field fixes the preferred time direction as well as the preferred frame of the phase space. 

Besides the construction of the theory, we have discussed the cosmological dynamics and the linear perturbations. The canonical transformation generically changes the speed of the gravitational waves. Hence, the canonical transformation should reduce to a trivial transformation in the late-time universe since the speed of the gravitational waves has to be the same as the speed of light with a high degree of accuracy ($\lesssim 10^{-15}$) at the present universe~\cite{TheLIGOScientific:2017qsa,Monitor:2017mdv}. On the other hand, there is no model-independent constraint on the speed of gravitational waves in the early universe. We have thus given an example of the theory representing the ultraviolet modification of gravity which yields a non-singular universe where the universe starts from the Minkowski spacetime.

Although we have considered a simple generating functional of the canonical transformation \eqref{trans}, one can consider more general canonical transformations. For instance, one may introduce the dependence of $f$ on $\pi^{ik}\pi^{jl}\gamma_{ij}\gamma_{kl}/{\rm det}\gamma $. Many new and yet unexplored theories of gravity can be generated from known theories via canonical transformations.

Our procedure to introduce the matter field in a consistent way can be applied to the minimally modified theories of gravity~\cite{Lin:2017oow}. As pointed out in~\cite{Carballo-Rubio:2018czn}, the matter coupling in the minimally modified theories of gravity is a nontrivial task. If a theory has a first class constraint, one may a priori introduce a gauge fixing condition just in the same way as in the present case. Then, the matter field can be consistently introduced. 

In summary, the present paper gives explicit
examples of gravity theories that have only 2 local physical
degrees of freedom, whose actions are written in terms of the
metric only (after integrating out the auxiliary field $\Phi$), to which matter fields can be consistently coupled,
and that have observable predictions different from GR. Although the present paper focuses on gravitational theories via the canonical transformation followed by a novel matter coupling, our formalism can be used for other gravitational theories with two local degrees of freedom.
Since we now have a new matter coupling, it would be interesting to investigate observational consequences. For example, the theories obtained by the canonical transformation must admit black hole solutions since the theory is equivalent to GR in vacuum. However, it is nontrivial how the black hole is formed and observed in such theories because the matter propagates on the frame that is related to the original Einstein frame by a non-trivial canonical transformation. We leave further investigations to a future work.

\section*{Acknowledgments}
The work of K.A. was supported in part by Grants-in-Aid from the Scientific Research Fund of the Japan Society for the Promotion of Science  (No. 15J05540). The work of S.M. was supported by Japan Society for the Promotion of Science (JSPS) Grants-in-Aid for Scientific Research (KAKENHI) No.\ 17H02890, No.\ 17H06359, and by World Premier International Research Center Initiative (WPI), MEXT, Japan..   The work of C.L. is carried out under POLONEZ programme of Polish National Science Centre, No. UMO-2016/23/P/ST2/04240,  which has received funding from the European Union's Horizon 2020 research and innovation programme under the Marie Sk\l odowska-Curie grant agreement NO. 665778.  \includegraphics[width=0.08\textwidth]{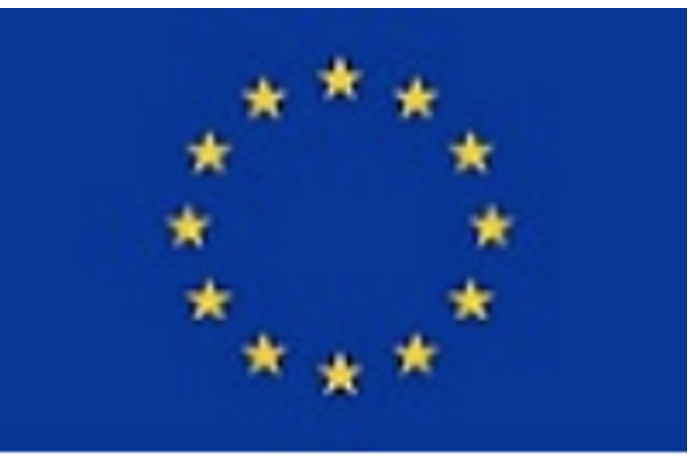}


\bibliography{ref}

\providecommand{\href}[2]{#2}\begingroup\raggedright\begin{thebibliography}{10}

\bibitem{Fujii:2003pa}
Y.~Fujii and K.~Maeda, \emph{{The scalar-tensor theory of gravitation}}.
\newblock Cambridge University Press, 2007.

\bibitem{Bekenstein:1992pj}
J.~D. Bekenstein, \emph{{The Relation between physical and gravitational
  geometry}}, \href{http://dx.doi.org/10.1103/PhysRevD.48.3641}{\emph{Phys.
  Rev.} {\bf D48} (1993) 3641--3647},
  [\href{http://arxiv.org/abs/gr-qc/9211017}{{\tt gr-qc/9211017}}].

\bibitem{Horndeski:1974wa}
G.~W. Horndeski, \emph{{Second-order scalar-tensor field equations in a
  four-dimensional space}},
  \href{http://dx.doi.org/10.1007/BF01807638}{\emph{Int. J. Theor. Phys.} {\bf
  10} (1974) 363--384}.

\bibitem{Charmousis:2011bf}
C.~Charmousis, E.~J. Copeland, A.~Padilla and P.~M. Saffin, \emph{{General
  second order scalar-tensor theory, self tuning, and the Fab Four}},
  \href{http://dx.doi.org/10.1103/PhysRevLett.108.051101}{\emph{Phys. Rev.
  Lett.} {\bf 108} (2012) 051101}, [\href{http://arxiv.org/abs/1106.2000}{{\tt
  1106.2000}}].

\bibitem{Deffayet:2011gz}
C.~Deffayet, X.~Gao, D.~A. Steer and G.~Zahariade, \emph{{From k-essence to
  generalised Galileons}},
  \href{http://dx.doi.org/10.1103/PhysRevD.84.064039}{\emph{Phys. Rev.} {\bf
  D84} (2011) 064039}, [\href{http://arxiv.org/abs/1103.3260}{{\tt
  1103.3260}}].

\bibitem{Kobayashi:2011nu}
T.~Kobayashi, M.~Yamaguchi and J.~Yokoyama, \emph{{Generalized G-inflation:
  Inflation with the most general second-order field equations}},
  \href{http://dx.doi.org/10.1143/PTP.126.511}{\emph{Prog. Theor. Phys.} {\bf
  126} (2011) 511--529}, [\href{http://arxiv.org/abs/1105.5723}{{\tt
  1105.5723}}].

\bibitem{Ezquiaga:2016nqo}
J.~M. Ezquiaga, J.~Garc{\'\i}a-Bellido and M.~Zumalac{\'a}rregui,
  \emph{{Towards the most general scalar-tensor theories of gravity: a unified
  approach in the language of differential forms}},
  \href{http://dx.doi.org/10.1103/PhysRevD.94.024005}{\emph{Phys. Rev.} {\bf
  D94} (2016) 024005}, [\href{http://arxiv.org/abs/1603.01269}{{\tt
  1603.01269}}].

\bibitem{Gleyzes:2014dya}
J.~Gleyzes, D.~Langlois, F.~Piazza and F.~Vernizzi, \emph{{Healthy theories
  beyond Horndeski}},
  \href{http://dx.doi.org/10.1103/PhysRevLett.114.211101}{\emph{Phys. Rev.
  Lett.} {\bf 114} (2015) 211101}, [\href{http://arxiv.org/abs/1404.6495}{{\tt
  1404.6495}}].

\bibitem{Langlois:2015cwa}
D.~Langlois and K.~Noui, \emph{{Degenerate higher derivative theories beyond
  Horndeski: evading the Ostrogradski instability}},
  \href{http://dx.doi.org/10.1088/1475-7516/2016/02/034}{\emph{JCAP} {\bf 1602}
  (2016) 034}, [\href{http://arxiv.org/abs/1510.06930}{{\tt 1510.06930}}].

\bibitem{Bettoni:2013diz}
D.~Bettoni and S.~Liberati, \emph{{Disformal invariance of second order
  scalar-tensor theories: Framing the Horndeski action}},
  \href{http://dx.doi.org/10.1103/PhysRevD.88.084020}{\emph{Phys. Rev.} {\bf
  D88} (2013) 084020}, [\href{http://arxiv.org/abs/1306.6724}{{\tt
  1306.6724}}].

\bibitem{Zumalacarregui:2013pma}
M.~Zumalac{\'a}rregui and J.~Garc{\'\i}a-Bellido, \emph{{Transforming gravity:
  from derivative couplings to matter to second-order scalar-tensor theories
  beyond the Horndeski Lagrangian}},
  \href{http://dx.doi.org/10.1103/PhysRevD.89.064046}{\emph{Phys. Rev.} {\bf
  D89} (2014) 064046}, [\href{http://arxiv.org/abs/1308.4685}{{\tt
  1308.4685}}].

\bibitem{Gleyzes:2014qga}
J.~Gleyzes, D.~Langlois, F.~Piazza and F.~Vernizzi, \emph{{Exploring
  gravitational theories beyond Horndeski}},
  \href{http://dx.doi.org/10.1088/1475-7516/2015/02/018}{\emph{JCAP} {\bf 1502}
  (2015) 018}, [\href{http://arxiv.org/abs/1408.1952}{{\tt 1408.1952}}].

\bibitem{Crisostomi:2016czh}
M.~Crisostomi, K.~Koyama and G.~Tasinato, \emph{{Extended Scalar-Tensor
  Theories of Gravity}},
  \href{http://dx.doi.org/10.1088/1475-7516/2016/04/044}{\emph{JCAP} {\bf 1604}
  (2016) 044}, [\href{http://arxiv.org/abs/1602.03119}{{\tt 1602.03119}}].

\bibitem{Achour:2016rkg}
J.~Ben~Achour, D.~Langlois and K.~Noui, \emph{{Degenerate higher order
  scalar-tensor theories beyond Horndeski and disformal transformations}},
  \href{http://dx.doi.org/10.1103/PhysRevD.93.124005}{\emph{Phys. Rev.} {\bf
  D93} (2016) 124005}, [\href{http://arxiv.org/abs/1602.08398}{{\tt
  1602.08398}}].

\bibitem{BenAchour:2016fzp}
J.~Ben~Achour, M.~Crisostomi, K.~Koyama, D.~Langlois, K.~Noui and G.~Tasinato,
  \emph{{Degenerate higher order scalar-tensor theories beyond Horndeski up to
  cubic order}}, \href{http://dx.doi.org/10.1007/JHEP12(2016)100}{\emph{JHEP}
  {\bf 12} (2016) 100}, [\href{http://arxiv.org/abs/1608.08135}{{\tt
  1608.08135}}].

\bibitem{Ezquiaga:2017ner}
J.~M. Ezquiaga, J.~Garc{\'\i}a-Bellido and M.~Zumalac{\'a}rregui, \emph{{Field
  redefinitions in theories beyond Einstein gravity using the language of
  differential forms}},
  \href{http://dx.doi.org/10.1103/PhysRevD.95.084039}{\emph{Phys. Rev.} {\bf
  D95} (2017) 084039}, [\href{http://arxiv.org/abs/1701.05476}{{\tt
  1701.05476}}].

\bibitem{Domenech:2015tca}
G.~Dom{\`e}nech, S.~Mukohyama, R.~Namba, A.~Naruko, R.~Saitou and Y.~Watanabe,
  \emph{{Derivative-dependent metric transformation and physical degrees of
  freedom}}, \href{http://dx.doi.org/10.1103/PhysRevD.92.084027}{\emph{Phys.
  Rev.} {\bf D92} (2015) 084027}, [\href{http://arxiv.org/abs/1507.05390}{{\tt
  1507.05390}}].

\bibitem{Lin:2017oow}
C.~Lin and S.~Mukohyama, \emph{{A Class of Minimally Modified Gravity
  Theories}},
  \href{http://dx.doi.org/10.1088/1475-7516/2017/10/033}{\emph{JCAP} {\bf 1710}
  (2017) 033}, [\href{http://arxiv.org/abs/1708.03757}{{\tt 1708.03757}}].

\bibitem{Carballo-Rubio:2018czn}
R.~Carballo-Rubio, F.~Di~Filippo and S.~Liberati, \emph{{Minimally modified
  theories of gravity: a playground for testing the uniqueness of general
  relativity}},  \href{http://arxiv.org/abs/1802.02537}{{\tt 1802.02537}}.

\bibitem{Henneaux:2009zb}
M.~Henneaux, A.~Kleinschmidt and G.~Lucena~G{\'o}mez, \emph{{A dynamical
  inconsistency of Horava gravity}},
  \href{http://dx.doi.org/10.1103/PhysRevD.81.064002}{\emph{Phys. Rev.} {\bf
  D81} (2010) 064002}, [\href{http://arxiv.org/abs/0912.0399}{{\tt
  0912.0399}}].

\bibitem{Gumrukcuoglu:2016jbh}
A.~E. G{\"u}mr{\"u}k{\c{c}}{\"u}o{\u{g}}lu, S.~Mukohyama and T.~P. Sotiriou,
  \emph{{Low energy ghosts and the Jeans' instability}},
  \href{http://dx.doi.org/10.1103/PhysRevD.94.064001}{\emph{Phys. Rev.} {\bf
  D94} (2016) 064001}, [\href{http://arxiv.org/abs/1606.00618}{{\tt
  1606.00618}}].

\bibitem{TheLIGOScientific:2017qsa}
{\scshape Virgo, LIGO Scientific} collaboration, B.~Abbott et~al.,
  \emph{{GW170817: Observation of Gravitational Waves from a Binary Neutron
  Star Inspiral}},
  \href{http://dx.doi.org/10.1103/PhysRevLett.119.161101}{\emph{Phys. Rev.
  Lett.} {\bf 119} (2017) 161101}, [\href{http://arxiv.org/abs/1710.05832}{{\tt
  1710.05832}}].

\bibitem{Monitor:2017mdv}
{\scshape Virgo, Fermi-GBM, INTEGRAL, LIGO Scientific} collaboration, B.~P.
  Abbott et~al., \emph{{Gravitational Waves and Gamma-rays from a Binary
  Neutron Star Merger: GW170817 and GRB 170817A}},
  \href{http://dx.doi.org/10.3847/2041-8213/aa920c}{\emph{Astrophys. J.} {\bf
  848} (2017) L13}, [\href{http://arxiv.org/abs/1710.05834}{{\tt 1710.05834}}].

\end{thebibliography}\endgroup
\bibliographystyle{JHEP}

\end{document}